%% file: root.tex
\documentclass[conference]{IEEEtran}
\IEEEoverridecommandlockouts
\usepackage{cite}
\usepackage{amsmath,amssymb,amsfonts}
\usepackage{algorithmic}
\usepackage{graphicx}
\usepackage{textcomp}
\usepackage{xcolor}

\usepackage{algorithm2e} 

\usepackage{url}

\usepackage{subcaption}

\def\BibTeX{{\rm B\kern-.05em{\sc i\kern-.025em b}\kern-.08em
    T\kern-.1667em\lower.7ex\hbox{E}\kern-.125emX}}

\begin{document}

\title{Smart Resource Management for Data Streaming using an Online Bin-packing Strategy
\thanks{
The HASTE Project (http://haste.research.it.uu.se/) is funded by the Swedish Foundation for Strategic Research (SSF) under award no. BD15-0008, and the eSSENCE strategic collaboration for eScience.
}
}


\author{
    \IEEEauthorblockN{Oliver Stein\IEEEauthorrefmark{1},
    Ben Blamey\IEEEauthorrefmark{2},
    Johan Karlsson\IEEEauthorrefmark{3},
    Alan Sabirsh\IEEEauthorrefmark{3},
    Ola Spjuth\IEEEauthorrefmark{4},
    Andreas Hellander\IEEEauthorrefmark{2},
    and Salman Toor\IEEEauthorrefmark{2}
    }
    \IEEEauthorblockA{
    \IEEEauthorrefmark{1} Department of Pharmaceutical Biosciences, Uppsala University, Sweden\\ 
    Email: \{Oliver.Stein, Ola.Spjuth\}@farmbio.uu.se\\
    \IEEEauthorrefmark{2} Department of Information Technology, Division of Scientific Computing, Uppsala University, Sweden\\ 
    Email: \{Ben.Blamey, Andreas.Hellander, Salman.Toor\}@it.uu.se\\
    \IEEEauthorrefmark{3} Discovery Sciences, Innovative Medicines, AstraZeneca, Gothenburg, Sweden\\ 
    Email: \{Johan.Karlsson1,Alan.Sabrish\}@astrazeneca.com\\
    }
}

\maketitle

\thispagestyle{plain}
\pagestyle{plain}

\begin{abstract}
Data stream processing frameworks provide reliable and efficient mechanisms for executing complex workflows over large datasets. 
A common challenge for the majority of currently available streaming frameworks is efficient utilization of resources. Most frameworks use static or semi-static settings for resource utilization that work well for established use cases but lead to  marginal improvements for unseen scenarios. Another pressing issue is the efficient processing of large individual objects such as images and matrices typical for scientific datasets. HarmonicIO has proven to be a good solution for streams of relatively large individual objects, as demonstrated in a benchmark comparison with the Spark and Kafka streaming frameworks. We here present an extension of the HarmonicIO framework based on the online bin-packing algorithm, to allow for efficient utilization of resources. Based on a real world use case from large-scale microscopy pipelines, we compare results of the new system to Spark's auto-scaling mechanism.
\end{abstract}



\section{Introduction}
\label{sec:introduction}

\input{inputs/intro.tex}

\section{Related Work}
\label{sec:related_work}
\input{inputs/related.tex}

\section{The HarmonicIO Streaming Framework}
\label{sec:harmonicio}
\input{inputs/harmonicio.tex}

\section{Online Bin-packing}
\label{sec:bin_packing}
\input{inputs/binpacking.tex}

\section{Framework Architecture}
\label{sec:architecture}
\input{inputs/architecture.tex}
\section{Results and Discussion}
\label{sec:results_discussion}
\input{inputs/results.tex}

\section{Conclusion and Future Directions}
\label{sec:conclusion}
\input{inputs/conclusion.tex}



\bibliographystyle{IEEEtran}
\bibliography{IEEEabrv,references}

\end{document}

%% file: inputs/intro.tex
Production-grade data stream processing frameworks such as Spark\footnote{https://spark.apache.org/}, Kafka\footnote{https://kafka.apache.org/} and Storm\footnote{https://storm.apache.org/} have enabled efficient, complex analysis on large datasets. 
These frameworks feature reliable transfer of the data, efficient execution based on multiple processing units, in- or out-of-order processing, and recovery from failures. These features are fundamental
to develop production-grade streaming applications, but are not themselves sufficient to guarantee efficient utilization of resources.  Indeed, with the popularity of public cloud infrastructures   
based on a pay-as-you-go model, the extended list of requirements both for the streaming frameworks and for the applications that run using these frameworks 
include efficient utilization of resources to reduce the cost of running applications, and rapid deployment of frameworks on different platforms. To achieve this, streaming frameworks need to be resource-aware in order to achieve the best possible resource utilization based on different scaling mechanisms. 


Moreover, frameworks need to be flexible enough in terms of management of in-homogeneous compute and storage resources since this allows for scaling processing units based on the best possible prices available on the public platforms. On the application side, one requirement is to design self-contained applications that can be deployed seamlessly on a variety of resources. To that end, different efforts have been made. The most popular self-contained application design scheme is the containerization approach. Based on Docker\footnote{https://www.docker.com/}, LXD\footnote{https://linuxcontainers.org/} or Singularity\footnote{https://singularity.lbl.gov/} container software, applications can easily be deployed on a variety of resources. 

For intelligent resource management, different machine learning approaches both from supervised and unsupervised learning have been extensively studied \cite{KRAWCZYK2017132}. However, it has been observed that in order to make supervised learning approaches effective, regular re-training is required to cope with evolving scenarios \cite{khoshkbarforoushha2015resource, doi:10.1137/1.9781611974010.98}. Unsupervised learning on the other hand needs longer time to provide reasonable estimates \cite{10.1007/978-3-319-91632-3_8}. Additionally, a large quantity of the published work in this area is based on synthetic datasets where real environment challenges are not very well covered. 

Another challenge which is not well covered in the data streaming framework domain is the efficient processing of large individual objects. Most of the currently available streaming frameworks focus on processing very large datasets composed of many small individual data objects \cite{7530084}. This maps well to the major use case of analysis of massive social media datastreams in which each individual object is a text message or a picture of small size 
(from tens of bytes to kilobytes). On the other hand, for typical  scientific datasets composed of e.g. matrices, images or large text files, the individual objects are often relatively large (ranging from kilo- to giga\-bytes). 

Motivated by applications in large-scale processing of microscopy images, we recently developed HarmonicIO, a data streaming framework geared towards efficient processing of streams composed of large individual objects\cite{hio}. Recent performance benchmark comparisons of HarmonicIO with the Spark and Kafka streaming frameworks illustrate the increased throughput that can be expected from 
HarmonicIO in that scenario \cite{blamey2018apache}. The architecture and the features of HarmonicIO  are further discussed in Section~\ref{sec:harmonicio}.

In this article, we extend the HarmonicIO framework with features for efficient and intelligent resource utilization by introducing an Intelligent Resource Manager (IRM) component. Our approach is based on online bin-packing, a lightweight and extremely efficient algorithm that lets us optimally utilize available compute resources. 
In contrast to solutions based on machine learning, the online bin-packing algorithm does not require training data and model fitting. Instead, we employ a run-time learning process that profiles the characteristics of the running workloads. The proposed IRM extension thus relies on bin-packing in order to schedule the workloads based on their run-time resource usage. 

To test the IRM extension we evaluate the system in an environment hosted in the SNIC science cloud, with tests based on both synthetic and use-case based workloads. We show a high degree of efficiency in resource scheduling from the bin-packing algorithm, and we highlight the noise from running the system in real environments as the error between scheduled versus measured resource utilization. 


Specifically, we make the following key contributions:
\begin{itemize}
\item Extend the HarmonicIO framework with efficient resource utilization using online bin-packing algorithm. 
\item Provide an extensive evaluation of the proposed IRM component.
\item Thoroughly compare the here proposed resource allocation mechanism and HarmonicIO with Spark streaming for a real-world scientific workload.
\end{itemize}
The article highlights the underlying challenges and proposes a well-designed solution for efficient resource utilization in the highly dynamic environment of streaming frameworks. Our experiments are based on real environment settings and the presented comparison with the Spark streaming shows the strength of proposed solution.  

The remainder of the article is organized as follows. Section~\ref{sec:related_work} reviews state-of-the-art approaches for the efficient resource utilization in streaming frameworks. Section~\ref{sec:harmonicio} explains the architecture and the features of HarmonicIO. The online bin-packing algorithm is covered in Section~\ref{sec:bin_packing}. Section~\ref{sec:architecture} explains the integration details of the proposed IRM component and the HarmonicIO framework. Results are presented in the Section~\ref{sec:results_discussion} and Section~\ref{sec:conclusion} summarize the article and outlines future research directions.

%% file: inputs/related.tex
Various approaches have been explored to address the challenge of efficient resource utilization. For example, a popular domain has been control theory, with previous work investigating how to use Kalman filters to minimize operational resource costs~\cite{GHANBARI2012104}, or to track CPU usage and accordingly update resource allocations for workloads as they vary~\cite{Kalyvianaki:2009:SSC:1555228.1555261}. The interesting feature of Kalman filters is the predictive  estimations of future behaviour, allowing the workloads to be captured increasingly accurately. The difficulty in applying the filters to resource scheduling lies in modeling the cost functions to minimize and the control system, in order to achieve their full potential.
In contrast to these works that targets bare-metal and VM environments, the solution proposed in this article targets resource scheduling based on containers under a data streaming setting. The main goal of adaptive resource utilization optimization remains the same.



Another interesting approach is to use overbooking, as proposed by \cite{tomas2014autonomic}. They designed a model based on overbooking  combined with risk assessment to maintain tolerable performance levels and at the same time keep a minimum level of resource utilization across multiple resources (CPU, memory, network I/O etc.). This reduced overestimation of resource requirements and helped server application collocation. In comparison our approach assigns the scheduling of computing resources to the streaming framework rather than the user having to provide information about the workloads.


Bin-packing has previously been used for scheduling workloads in cloud computing contexts. In \cite{song2014adaptive}, a resource manager for cloud centres was proposed, featuring dynamic application reallocation with bin-packing based on run-time workloads. With the help of the resource manager, the number of VMs required to host the applications could be reduced. Based on these promising results, the work on our proposal was inspired by the use of bin-packing for optimizing resource utilization. However, we opted to use the bin-packing on a container level, gearing towards the very popular containerized approach today.

Furthermore, reinforcement learning (RL) is another appealing domain for exploring optimal auto-scaling policies. The methods rely on an exploration and exploitation approach. The idea is to get a reward or penalty based on the decision and a policy will be designed by maximizing the rewards. The method works very well in many dynamic systems. However, the challenge is to calibrate a trade-off between exploration and exploitation. With a strict exploitation approach the system will be reluctant to try new polices. On the other hand, too much exploration leads to longer time to set a policy. The paper \cite{Tesauro2007} discusses the marginal performance of the Q-learning RL method for auto-scaling scenarios. Furthermore, papers \cite{Li:2018:MCD:3199517.3199521, CARDELLINI2018171} present advanced RL approaches for auto-scaling scenarios. 
Our proposed approach based on bin-packing is not limited by the incentive based strategies, yet it is still flexible enough to adapt according to the dynamic requirements. 

%% file: inputs/harmonicio.tex
One of the components of the HASTE platform is the stream processing framework HarmonicIO (HIO), introduced in \cite{hio}. HIO uses Docker containers as \emph{processing engines} (PEs) to process the streamed messages; these are designed and provided by the client based on a template. A deployment of the framework consists of a single master node, and several worker nodes hosting PE containers. 

To work with HIO, a user will first need to design their desired data processing task as a Docker container image and publish this to Docker Hub; examples and a template are available at our GitHub repository\footnote{https://github.com/HASTE-project/HarmonicPE}. With this in place the user can start querying a HIO server to host PEs based on this container image and the stream data to process, using the \emph{HarmonicIO Stream Connector}, a Python client API. From here HIO takes care of directing the data streams to the intended PE endpoint. Figure~\ref{fig:hio_concept} is an illustration of the architecture. A key feature is smart P2P behaviour, where messages are forwarded directly to available PEs for processing, falling back to a backlog queue on the master node as needed.


One of the strengths of HIO lies in the throughput for larger object sizes. Since one of the goals of the HASTE project is to analyze microscopy images, the target object size is much different to the typical workloads often consisting of streaming text files or JSON objects. \cite{blamey2018apache} compared HIO to similar common streaming frameworks, namely Apache Spark Streaming and Kafka, and found that HIO could achieve higher throughput for relevant object sizes, under some configurations.

In \cite{thesis_irm}, several changes were made to HIO in order to add support for dynamic workload scheduling and improve resource management. The main contribution was to add the possibility for the master node to autonomously decide where to host containers and remove any user-to-worker communication, making way for an algorithm that could optimize this decision. The outcome of the work led to the extension of HIO proposed in this article which adds intelligent resource management based on bin-packing, presented briefly in Section~\ref{sec:system}.




\subsection{Architecture}
\label{subsec:hio_components}



HIO's peer-to-peer (P2P) architecture allows messages to be transferred directly from source nodes to workers for processing, falling back to a queue at the master node if processing capacity exceeds message ingress. Messages in this queue are processed with higher priority than new messages. 

\begin{figure}[tbp]
    \begin{center}
    \includegraphics[width=\linewidth]{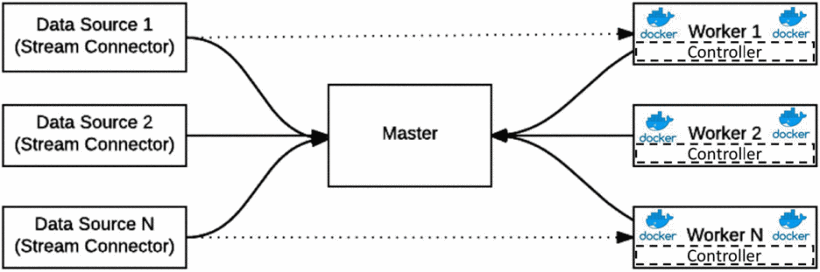}
    \end{center}
    \caption{The HarmonicIO architecture. The system consists of a master node, worker nodes and stream connectors, where solid lines indicate communication and dotted lines P2P data transfer. Image from \cite[Fig. 1]{hio}. 
    }
    \label{fig:hio_concept}
\end{figure}

HarmonicIO has the following overall architecture (see Figure~\ref{fig:hio_concept}): 
 
\begin{itemize}
\item \textbf{Stream connector} The stream connector acts as  the client to the HIO platform, handling communication with the REST APIs of the other nodes, so that the user can stream a message. Internally, it requests the address of an available PE, so the message can be sent directly if possible.
A stream request message consists of both the data to be processed, and the docker container and tag that a PE needs to run to process the data.
\item \textbf{Master} The master node of HIO is responsible for maintaining the state of the system, tracking worker nodes, and the availability of their containers, connects stream requests to workers that are available and starting containers, or \textit{Processing Engines} (PEs) as per user requests. It also maintains a backlog queue of messages, if message influx exceeds available processing capacity.
\item \textbf{Worker} The workers host PEs which contain the user's code to process data that is streamed to the PE via P2P from the stream connectors. The workers nodes report to the Master node, and can be seen as a pool or set of available streaming endpoints. 
\end{itemize}

%% file: inputs/binpacking.tex
In \cite{Seiden:2002:OBP:585265.585269} \textit{bin-packing} algorithms are described as algorithms aimed at solving the optimization problem of assigning a sequence of items of fixed \textit{size} into a number of \textit{bins} using as few of these as possible. Furthermore,   \cite{coffman1983dynamic} describes  \textit{online} bin-packing as the case where each item in the input sequence is assigned one by one without knowledge about the following items, meaning that information about future items is not contributing to the placement.  \cite{Seiden:2002:OBP:585265.585269} also mentions the \textit{asymptotic performance ratio}, denoted $R$, which indicates the number of bins an algorithm needs as a factor of the number of bins in the optimal solution.
Denoting the optimal number as $O$, online bin-packing algorithms will thus use $RO$ containers. Several studies~\cite{epstein2012comparing,johnson1974worst,Lee:1985:SOB:3828.3833,gambosi2000algorithms} have analyzed the performance of these algorithms, and generally they perform well when comparing the cheap cost in time and memory to the approximation results.

\subsection{The First-Fit algorithm}
\label{subsubsec:anyfit}
Several online bin-packing algorithms were studied in \cite{epstein2012comparing}. In particular, they looked at a group of such algorithms that they call the Any-Fit group. Relatively simple, they share a common approach for finding the bins in which to put the next item, and the best performance ratio in the group is proven to be $R=1.7$. The common approach is detailed in Algorithm~\ref{alg:anyfit}, where the input is a sequence of items, $L=(a_1, a_2, \hdots , a_n),\ a_i \in (0,1]$. The items are packed in order and $a_i$ corresponds to the item size. The list of currently active bins is denoted as $B = (b_1, b_2, \hdots, b_m)$, and $m$ is the number of bins needed at the end of the algorithm. As indicated, new bins are only generated when no currently active bin can fit the next item.

Of particular interest is the First-Fit algorithm, with a ratio of $R=1.7$ as well as  $O(n\ \mathrm{log}\ n)$-time and $O(n)$-space complexities, and is the algorithm that we based our resource management optimization upon. The search criterion in First-Fit is to find the first (lowest index) available bin in the list in which the current item fits.


\begin{algorithm}
\caption{General Any-Fit approach}
\label{alg:anyfit}
	\For{$i:=1$ \KwTo $n$}{
    	\Begin{
            find available bin $b_{\mathrm{a}}$ in $B$ according to criterion\\
            \eIf{$a_i$ fits in $b_{\mathrm{a}}$}{
                place $a_i$ in bin $b_{\mathrm{a}}$
            }{
                allocate new bin $b_{\mathrm{new}}$ and add to $B$\\
                place $a_i$ in bin $b_{\mathrm{new}}$
            }
    	}
	}
\end{algorithm}



%% file: inputs/architecture.tex
\label{sec:system}
In previous work~\cite{thesis_irm}, the \textit{Intelligent Resource Manager}  (IRM) system was designed as an extension of HIO based on the First-Fit bin-packing algorithm described above. Here, an overview of the bin-packing implementation and of the architecture of the extension components is presented.

\subsection{Resource management with bin-packing}
\label{subsec:method}
In order to improve the resource utilization, the PE containers are scheduled with the help of bin-packing. Modelling PEs as bin items, workers as bins and the workload resource usage as the item size, in theory the algorithm can provide the optimal way to schedule the containers in order to keep the number of workers needed down while not congesting resources. Thus the IRM continuously performs a bin-packing run on the currently waiting PEs.


Based on the bin-packing result, HIO can determine where to host the containers and in addition whether more or fewer worker nodes are needed for the current workload autonomously. This way auto-scaling of worker nodes is achieved. As for auto-scaling of the PE containers, the IRM looks at the rate of change in the streaming message queue length to evaluate whether HIO is consuming stream requests at a high enough rate. If not, the IRM will queue more PEs in order to drive down the waiting time for stream requests. After a time of being idle, a PE will self-terminate gracefully in order to free the resources.

In \cite{thesis_irm} the main metric for resource usage is the \textit{average CPU usage}, which is measured as a sliding time window average. The average usage is directly used as the item size for the bin-packing algorithm. Furthermore, in order to not block the system when the workload pressure increases, a small buffer of idle workers are kept ready to accept stream requests. This buffer is logarithmically proportional to the number of currently active workers, providing more headroom for fluctuations when the workload is not as high.

\subsection{IRM architecture}

\begin{figure}[ht]
\begin{center}
\includegraphics[width=\linewidth]{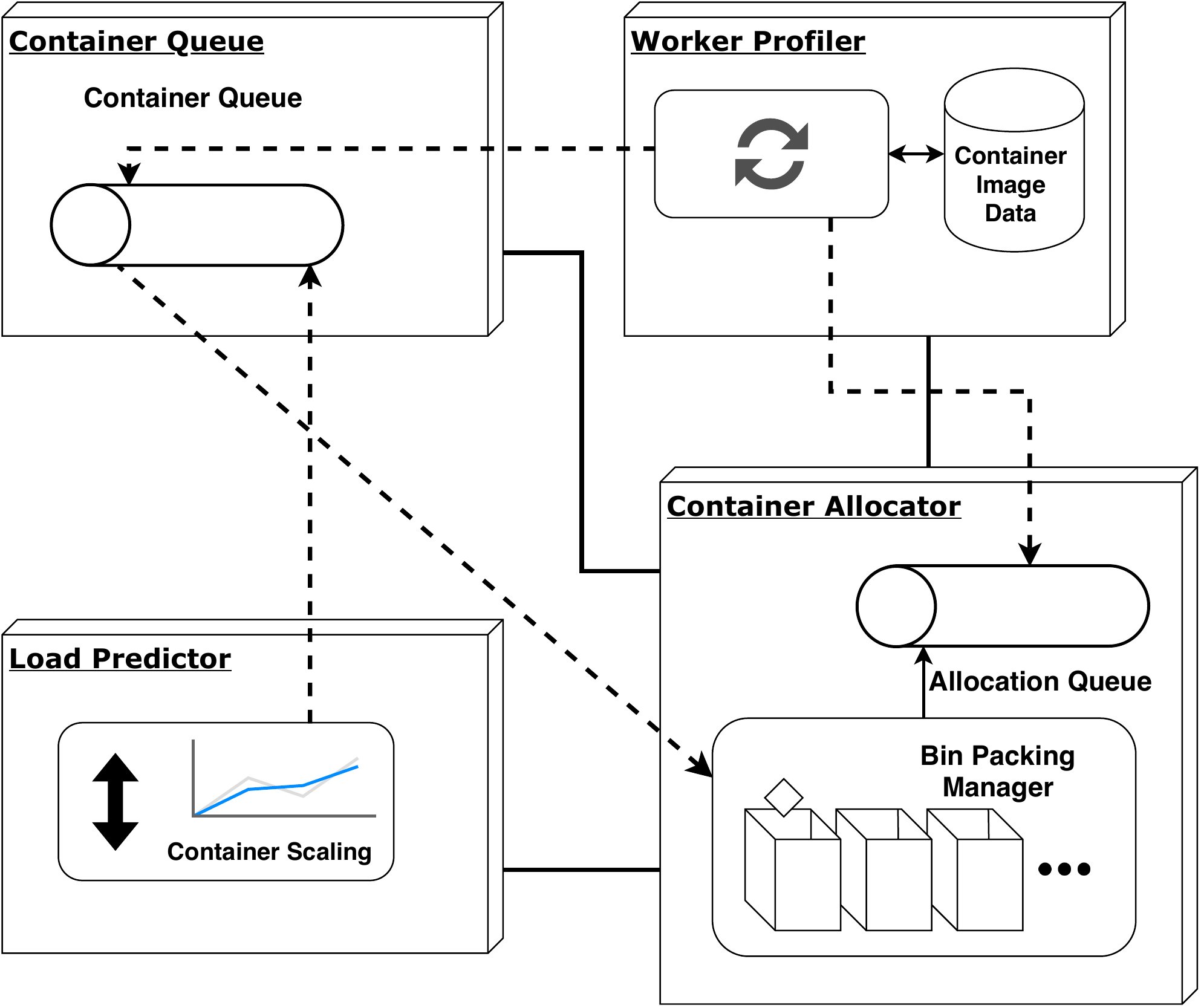}
\end{center}
\caption{Architecture overview of the IRM subcomponents. In particular the communication is drawn with dashed lines, showing the direction of data and communication. Figure from \cite[Fig. 2]{thesis_irm}.}
\label{fig:irm_schematic}
\end{figure}

In Figure~\ref{fig:irm_schematic}, the architecture of the IRM extension is illustrated, showing the four main components of the system; the \textbf{container queue}, \textbf{container allocator}, \textbf{load predictor} and \textbf{worker profiler}. The following sections detail these components further. As described in further detail in \cite{thesis_irm}, particularly in Section~4.3 and Table 1, there are many configurable parameters that control the behaviour of the IRM extension, which are briefly mentioned where relevant here.

\subsubsection{Container queue}
Whenever a PE is to be created, it must first enter the \textit{container queue}, a FIFO queue of container hosting requests. Each request contains the container image name, a time-to-live (TTL) counter, any metrics related to that image etc. The TTL counter is used in case the request is re-queued following a failed hosting attempt.

While waiting in the queue, requests are periodically updated with metric changes and finally consumed and processed by the periodic bin-packing algorithm. The queue holds requests both from auto-scaling decisions and manual hosting requests from users.



\subsubsection{Container allocator}
The \textit{container allocator} includes the bin-packing manager, responsible for performing the chosen bin-packing algorithm, and the allocation queue. As mentioned in Section~\ref{subsec:method}, in this model a worker VM represents a bin and the container hosting requests represent items. Active VMs indicate open bins, i.e. the VM is enabled as a host for PEs, with a capacity of $1.0$. The container requests have item sizes in the range $(0,1]$, indicating the CPU usage of that PE from $0-100\%$.

The bin-packing manager performs a bin-packing run at a configurable rate based on this representation, resulting in a mapping of where to host the queued PEs and how many worker WMs are needed to host these. The destination worker is attached to each container host request which is forwarded to the allocation queue. As requests are consumed the allocator attempts to start the PEs on the destination worker. In case a PE could not be started, for example if the target worker is a new VM still initializing, any information related to the target worker is removed from the request and it is sent to the container queue.

\subsubsection{Worker profiler}
To understand the resource utilization characteristics, the \textit{worker profiler} gathers runtime statistics from the workloads and is designed in two parts; the first part lies within the worker VMs, periodically measuring the current CPU usage for each running PE. The average per container image is calculated and sent to the master VM. The second part in turn aggregates the information from all active workers and keeps a moving average of the CPU utilization based on the last $N$ measurements, $N$ being arbitrarily configurable. Based on this information, the worker profiler provides an idea of the CPU utilization for PE container images that have been hosted on HIO previously. The average CPU is used by the bin-packing manager as the item size and the updated averages are propagated to container requests in the container and allocation queues.




\subsubsection{Load predictor}
The \textit{load predictor} is responsible for tracking the pressure of the streaming requests to HIO. Looking at the length of the message queue and its rate of change (ROC), the load predictor can determine if the rate of processing data streams is too slow and there is a need to add more PEs. The ROC provides  predictions for the need to scale up, and scheduling PEs this way gives HIO time to set up additional workers and reduces the congestion.

The decision of scaling up is based on various thresholds of the message queue length and ROC. These thresholds are configurable, and there are four cases, resulting in either a large or small increase in PEs. In short, if the ROC is very large or the queue is very long, this indicates that data streams are not processed fast enough. Reading the queue metrics is done periodically, and there is a timeout period after scheduling more PEs before the load predictor can do this again.

%% file: inputs/results.tex
Experiments on the IRM-extended HIO system has shown promising results. As part of the thesis project \cite{thesis_irm}, tests based on synthetic workloads were performed to evaluate the performance of the bin-packing implementation and the effect on resource utilization in HIO. The main outcome of these experiments are summarized in Section~\ref{sec:thesis_tests}.
Furthermore, new experiments have been conducted for a real world image-analysis use-case in collaboration with AstraZeneca. 
The results of these experiments are presented in Section~\ref{sec:haste_usecase}.

\subsection{IRM evaluation  experiments on synthetic workloads}
\label{sec:thesis_tests}
The IRM was tasked with profiling and scheduling workloads based on busying the CPU for specified usage levels and durations, mimicking a scenario where the bin-packing manager deals with items of various sizes and durations.

The main scenario that was experimented with included four different workloads all targeting $100\%$ CPU utilization for various amounts of time. These were streamed in regular small batches of jobs and two peaks of large batches to introduce different levels of intensity in pressure to the IRM. Some of the results from these experiments are shown and briefly discussed here.

\subsubsection{Efficient Resource Utilization}
\label{sec:simulation_resource_utilization}
\begin{figure}
    \centering
    \includegraphics[width=\linewidth]{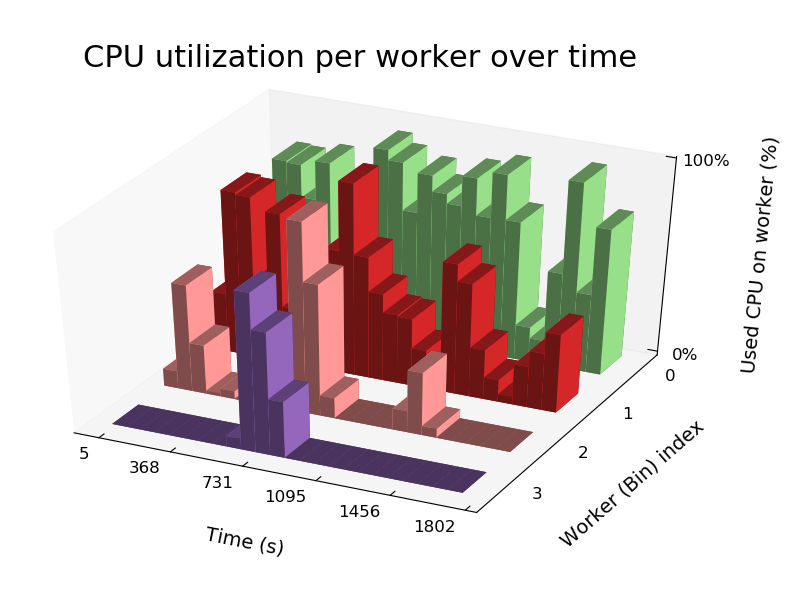}
    \caption{The CPU usage from $0-100\%$ per worker over time in 3D. Image reused from \cite[Fig. 8]{thesis_irm}.}    
    \label{fig:ref_run_3d}
\end{figure}

The results of the experiments indicate that the resource utilization may indeed benefit from the bin-packing approach. In Figure~\ref{fig:ref_run_3d}, the CPU usage per worker over time is shown in 3D-plots, giving an overview of the distribution of the jobs over the workers throughout the experiment. It is clear the workload is focused toward the lower index workers, leaving windows of time during which the higher index bins could be deactivated and the resources freed. 

%

\begin{figure}
    \centering
    \includegraphics[width=\linewidth]{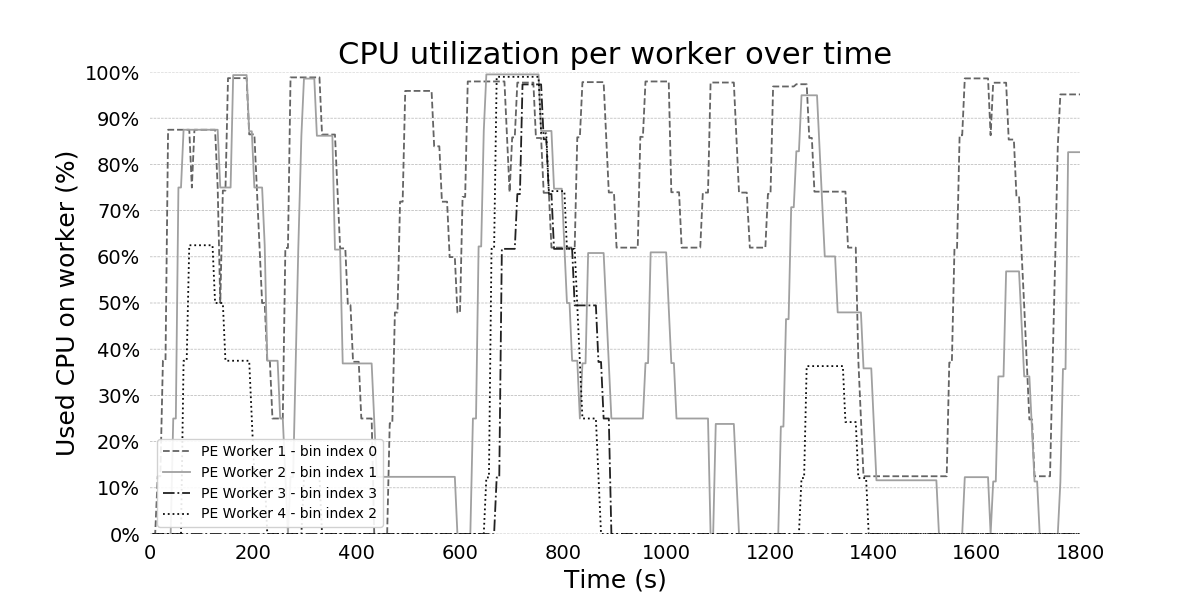}
    \caption{Scheduled CPU usage per worker over time from the bin-packing manager. Plot data from \cite[Fig. 9]{thesis_irm}.}
    \label{fig:ref_run_2d}
\end{figure}

Figure~\ref{fig:ref_run_2d} shows the CPU utilization over time per worker as a 2D-plot, giving a better view of the utilization levels. In general, the utilization of the workers peak at between $90-100\%$ CPU usage. At this point, a worker can not fit further jobs and any following workloads are scheduled to a higher index worker, following the expected behaviour of the bin-packing algorithm.

\subsubsection{Algorithm's Accuracy and Performance}
\label{sec:simulation_performance}
\begin{figure}
    \centering
    \includegraphics[width=\linewidth]{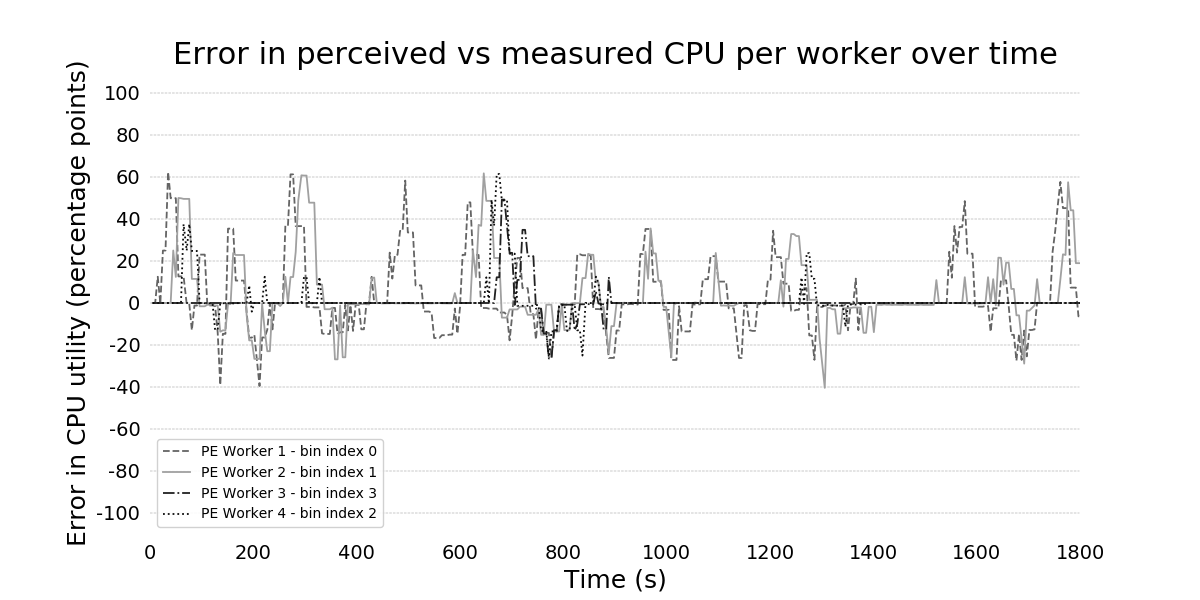}
    \caption{Error in percentage points between CPU usage scheduled by the bin-packing manager vs measured CPU usage over time per worker. Plot data from \cite[Fig. 10]{thesis_irm}.}
    \label{fig:ref_run_error}
\end{figure}

Figure~\ref{fig:ref_run_error} shows a plot of the error between the CPU usage as scheduled by the bin-packing manager and the measured CPU usage for each worker over time. The plot has a high amount of noise which is mainly due to the delay in starting and stopping containers compared to when they are scheduled to start and stop. This is discussed further in \cite{thesis_irm}. Another contributing factor is the irregularity in how often workloads are streamed, likely leading to PEs often starting and finishing.


With the error plot in mind, it is hard to judge the real CPU utilization efficiency that is achieved with the IRM extension for this experiment. However, the test scenario was also designed to stream images at varying frequency, which impacts the ability of the framework to keep a constant high efficiency. 

Nonetheless, the results indicate overall that bin-packing provides an appealing approach that is able to efficiently schedule containers in cloud computing contexts. Furthermore, the impact of the noisy error is hard to determine in a scenario with synthetic workloads and in heterogeneous distributed settings. More data from scenarios based on real use cases may help to better understand and further improve the framework.

\subsection{Image streams from quantitative microscopy}
\label{sec:haste_usecase}
\begin{figure}
    \centering
    \includegraphics[width=0.7\columnwidth]{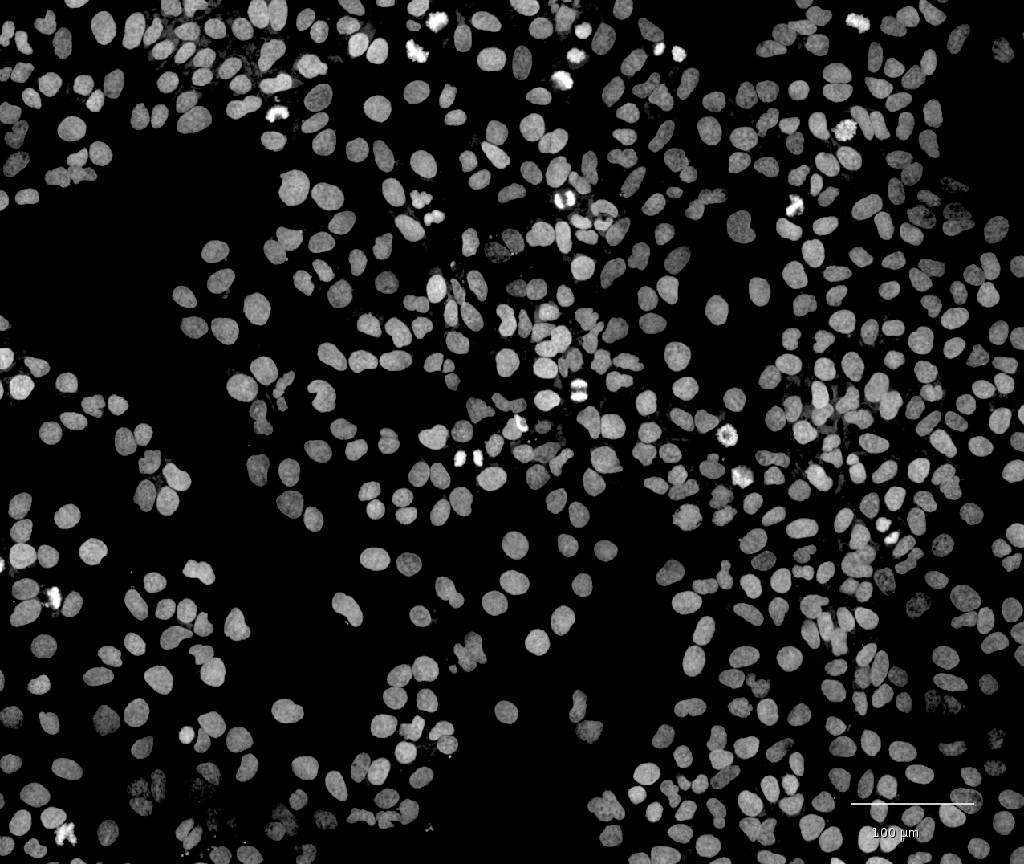}
    \caption{Shown is a representative image of Huh7 liver carcinoma cells seeded at a moderate density, stained with the cell-permeable DNA binding dye Hoechst 33342, and imaged using a confocal microscope. The dye binds primarily to the DNA in the cell nucleus resulting in a fluorescence image that shows only the cell nuclei and not the cytoplasm or cell membranes. The scale bar shows $100\mu m$.}
    \label{fig:Huh7}
\end{figure}

The data provided by AstraZeneca consists of a set of microscopy images (2 fields of view per well obtained using a Yokogawa CV7000 robotic confocal microscope fitted with a 20X Olympus air objective). Huh-7 cells (a hepatocellular lineage seeded at 6 different densities across a Perkin Elmer CellCarrier 384 well plate one day prior to imaging) were stained with nuclear dye prior to imaging (Hoechst 33342, ThermoFisher), and a CellProfiler analysis pipeline (Windows release 3.1.9) was created to count the number of nuclei and measure their areas. Due to variations in the images they take varying amounts of time to process, and the dataset includes a total of 767 images. Figure~\ref{fig:Huh7} shows an image from the dataset. A central future goal is to develop intelligent workflows that allows for dynamic and online re-configuration and data prioritization for large-scale and long-running imaging experiments. For that, high-throughput stream  analysis of images will be necessary.      

This CellProfiler pipeline was adapted to run within containers designed to be hosted in both HarmonicIO and Apache Spark Streaming environments. The next sections describe the experiments and results of running the image analysis in the two systems.

\subsubsection{Apache Spark Streaming}

For comparison with the system discussed, an Apache Spark Streaming application was developed and benchmarked for an equivalent image processing task. With the Spark File Streaming source, for each new image, CellProfiler is invoked as an external process to analyze the image, using a pre-configured image processing pipeline.

The use case is perhaps a little ill-suited to Spark: for one, the file names need to be passed to CellProfiler, but they are not easily available from the Spark APIs via the HDFS driver. Since our application is atypical, some work was needed to achieve satisfactory auto-scaling behaviour. There is support for \emph{dynamic allocation} (that is, scaling the Spark application within the cluster) specifically for  streaming, since Spark 2.0.0, (configured with the settings \texttt{spark.streaming.dynamicAllocation.*}), taking into consideration the batch processing and addressing other issues (\url{https://issues.apache.org/jira/browse/SPARK-12133}). However, our initial attempts to achieve satisfactory auto-scaling with this approach were problematic, because it begins to scale up only when a batch is completed. So, when the system is initially idle (with a single executor), the initial set of images for a 5 second batch interval (50 or more), each with having an execution time of 10-20 seconds, meant that the first batch takes minutes to execute, leaving the other available cores in the cluster unused. 
Because the images are processed by invoking CellProfiler as an external process, it is the minimum unit of parallelism. 

For this reason, we resorted to using the older \emph{dynamic allocation} functionality, with an \texttt{spark\allowbreak.dynamicAllocation\allowbreak.executorIdleTimeout} of 20 seconds. This begins scaling during the first batch. We also raised the \texttt{spark.streaming\allowbreak.concurrentJobs} setting from 1 to 3, so that other cores could begin processing the next batch while waiting for the `tail' of images within the job to each finish their 10-20 seconds of processing. This gave satisfactory auto-scaling performance.

The application was written in Scala, and is available online (including all the scaling settings)\footnote{github.com/HASTE-project/bin-packing-paper/blob/master/spark/spark-scala-cellprofiler/src/main/scala/CellProfilerStreaming.scala}.
The source directory was mounted as an NFS share on all the machines in the Spark cluster, because the image sizes (order MB) are too small to warrant the overhead of HDFS. CellProfiler and its dependencies were installed on all the machines in the Spark cluster.

\begin{figure}
    \centering
    \includegraphics[width=\columnwidth]{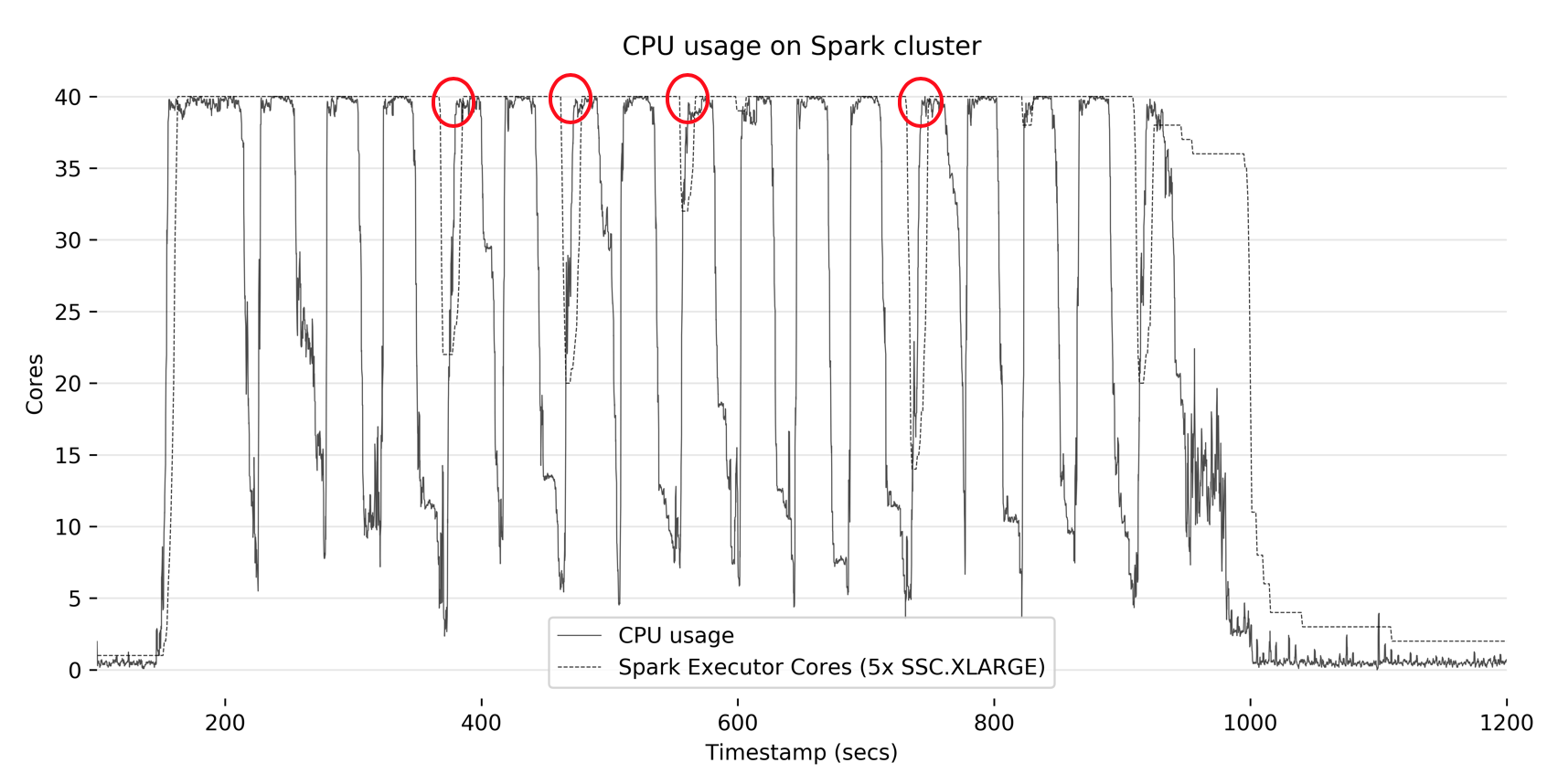}
    \caption{Spark executor cores vs. actual CPU usage. Red circles show where the Spark application was scaled down.}
    \label{fig:spark}
\end{figure}

Spark Version 2.3.0 was used. The cluster was deployed to SNIC, a cloud computing service for Swedish academia. The cluster consisted of 1xSSC.xlarge (for the spark master), 5xSSC.xlarge (for the workers), and 1xSSC.small for the virtual machine hosting the images. For the benchmarking, the elapsed system CPU usage on all the workers was polled with \texttt{top} (we take the sum of user and kernel CPU time), and the number of executor cores was polled via the Spark REST API. The clocks on all machines were initially synchronised with \texttt{ntpdate}. By combining the log files, Figure~\ref{fig:spark} was generated showing the real CPU used (shown number of cores) and the total cluster executor cores reported by the REST API. 

A number of phenomenon are visible in Figure~\ref{fig:spark}. We clearly see the executor cores scale up and down for the batch processing. The system scales to use all the available 40 worker cores in the cluster. If we look closely, we can see that the CPU usage leads the available cores by a few seconds when scaling up, as the executors get started, which is to be expected.

For unknown reasons, the system sat idle with 2 executors for some time. Secondly, despite increasing the number of concurrent jobs, we can clearly see each batch in the actual CPU usage, with idle gaps in between. In some cases, these gaps are sufficiently long that the system scales down 
(shown with red circles).
It is unclear why this is so, as only minimal data was collected back to the client application. The time could have been spent reading the images from disk. Profiling of the network usage would confirm this. 

\subsubsection{HarmonicIO with IRM}
For testing the image analysis pipeline in HIO 
the HIO stream connector was used to stream the entire collection of images as a single batch. For each image, CellProfiler is invoked with the single image located in a folder used as input. 

The setup was similar to the Spark experiment. The VMs for HIO were deployed on the SNIC science cloud \cite{ssc}, with one master node (SSC.xlarge), five worker nodes (SSC.xlarge) and one client (SSC.large). The IRM configuration for HIO uses the same default values presented in \cite{thesis_irm}, with the additional worker parameters \texttt{report\_interval} and \texttt{container\_idle\_timeout} both at 1 second.

In total, 10 runs of the experiment scenario were conducted, during which time data was gathered from the HIO system.
For each run, the streaming order of the images was randomized. HIO was started fresh for the first run and remained running for all subsequent runs. This allows us to evaluate the process of profiling the workloads. As expected, the initial run performed slightly worse than subsequent runs. This is because the initial guess for the new workload gets adjusted as the IRM gets a better profile of the CPU usage of the workload. From the second run and onward, the results differ only marginally, mainly due to the randomized streaming order. All figures represent the 10th and final run unless otherwise noted.

\begin{figure}[h]
    \centering
    \includegraphics[width=\linewidth]{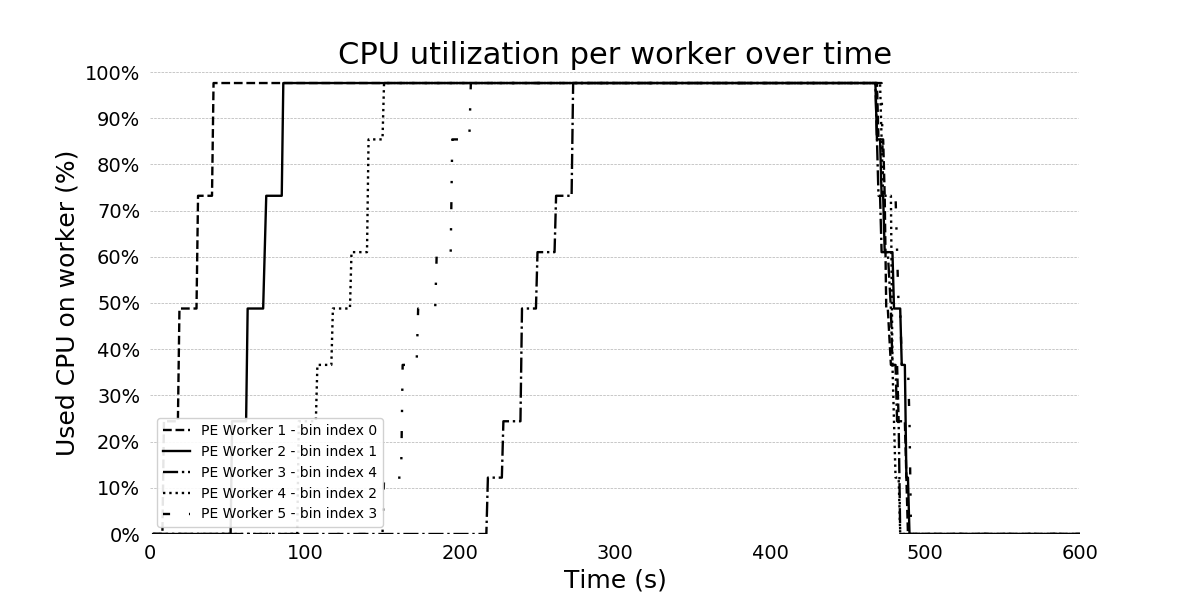}
    \caption{Bin-packing scheduled CPU usage per worker over time.}
    \label{fig:cellprofiler_cpu_run10}
\end{figure}
\begin{figure}[h]
    \centering
    \includegraphics[width=\linewidth]{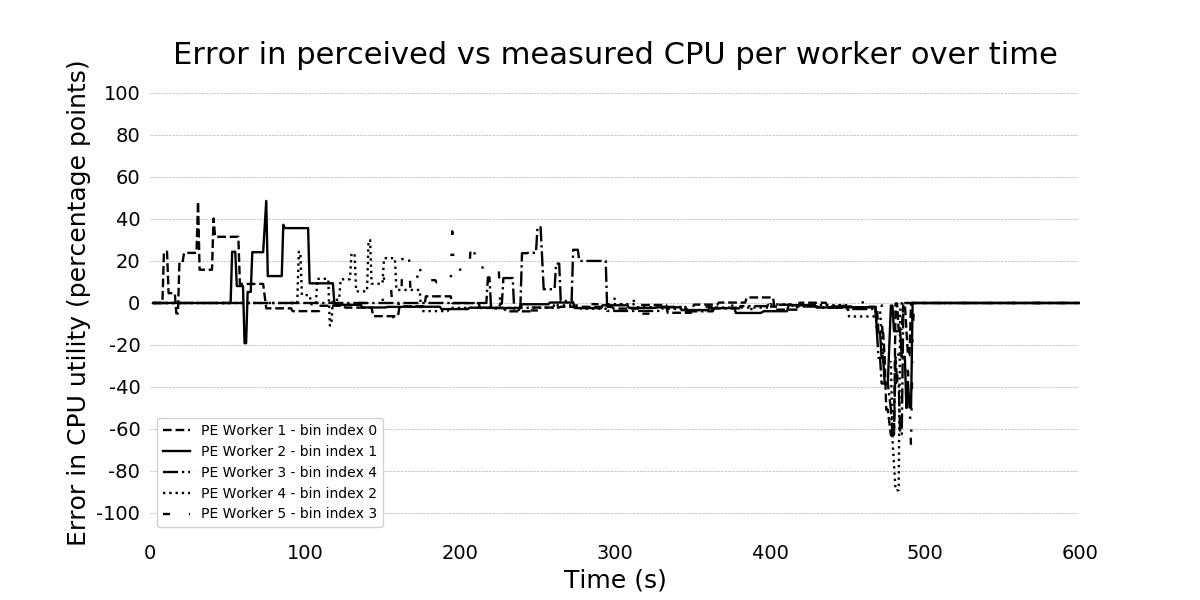}
    \caption{Error between perceived CPU and measured CPU usage in percentage points.}
    \label{fig:cellprofiler_error_run10}
\end{figure}

Starting with the CPU utilization, Figure~\ref{fig:cellprofiler_cpu_run10} illustrates how the bin-packing manager scheduled the CPU usage across the workers throughout the run. As visible in the plot, the workers are scheduled to use nearly $100\%$ CPU usage before the auto-scaling drives workloads to the next worker. The next plot in Figure~\ref{fig:cellprofiler_error_run10} shows the error between the scheduled CPU usage as perceived by the bin-packing manager and the measured CPU usage per worker. 

If we compare the plots in Figure~\ref{fig:cellprofiler_cpu_run10} and Figure~\ref{fig:cellprofiler_error_run10}, we can see that the bumps in the error for each worker coincide with the periods during which the bin-packing manager increases the number of PEs on that worker. Building on the hypothesis from the errors in the synthetic workload tests, this is in line with the expectations. Since the PEs will take a few moments to start processing incoming data after having been scheduled, there will be a difference between scheduled CPU and measured CPU usage. After this period the error settles close to 0, indicating that the scheduled workloads are quite accurate and the CPU utilization is close to ideal in terms of the bin-packing scheduling. The sudden large decrease in the error is the inverse, where the containers start shutting down from being idle in rapid succession.

Thus, we argue that despite the noise in the error plot due to startup and shutdown time of containers, the bin-packing algorithm is evidently efficient in the scheduling of CPU resources. Basing the IRM on bin-packing we manage to schedule the resources close to $100\%$ utilization levels.

\begin{figure}
    \centering
    \includegraphics[width=\linewidth]{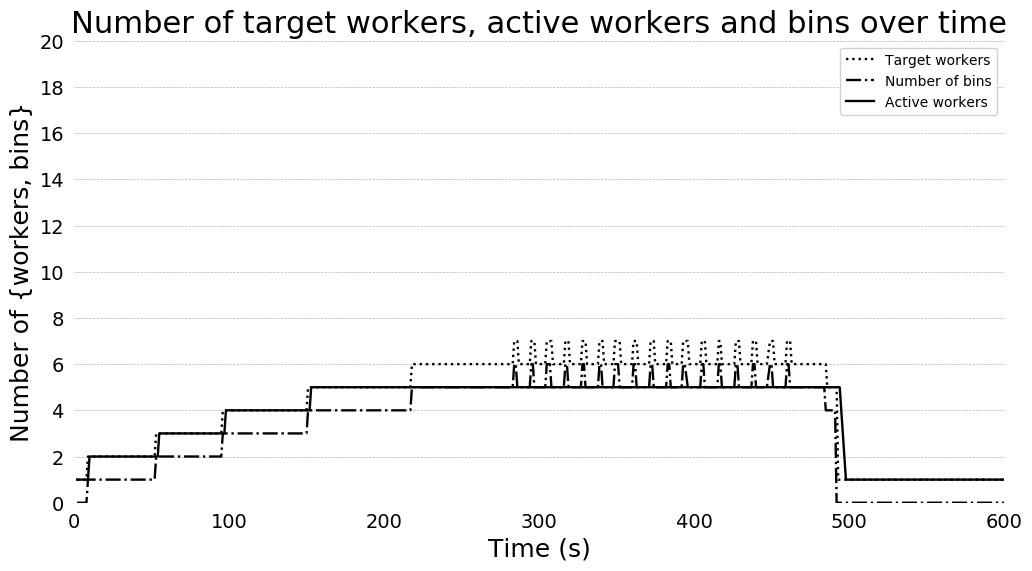}
    \caption{Number of target and current workers, and number of active bins, over time.}
    \label{fig:cellprofiler_workers_run10}
\end{figure}

The number of active workers as opposed to the target number and the ideal number of bins is plotted in Figure~\ref{fig:cellprofiler_workers_run10}. This plot illustrates that the IRM would have scheduled more workers to handle the workload if they were available. In order to make a fair comparison of the efficient utilization of available resources with the Spark streaming framework, we have restricted both of the frameworks to 5 workers. The periodic attempts to increase further are due to the IRM attempting to scale up, scheduling more PEs than can fit on the available 5 workers. These attempts will fail and the IRM constantly tries again to scale up until the queued images are processed. The plot in Figure~\ref{fig:cellprofiler_workers_run10} illustrates the efficient utilization of the available resources and at the same time IRM\textquotesingle s approach to actively look for more resources if possible.

An interesting observation is that based on the timestamps (x-axis) available in Figure~\ref{fig:spark} and Figure~\ref{fig:cellprofiler_cpu_run10}, it is evident that HIO outperforms Spark in terms of the overall processing time for the workload. The execution time of the entire batch of images is nearly halved, however as noted the images could not be sent as a single batch in the Spark experiment. This reinforces that HarmonicIO\textquotesingle s strategy based on online bin-packing seems well suited for stream processing environments, and that the framework is better equipped for this kind of application. 


It is also interesting to note that judging by the plots, the performance of the IRM has been much better during the experiments with the real usecase compared to the experiments with synthetic workloads discussed in Section~\ref{sec:thesis_tests}. The main difference between the two cases was that all the images were streamed in a singular large batch as opposed to periodic small batches. Thus HIO was allowed to really push the CPU usage to it's intended level with constant processing.

%% file: inputs/conclusion.tex
The efficient management of resource utilization in streaming frameworks is a non-trivial task. The current streaming frameworks already have multiple parameters to tune to provide reliable processing of large datasets, and efficient resource utilization further adds to the complexity of the frameworks. It is nonetheless an important problem and must be addressed. More effort is needed both from industry and academia to explore methods of resource utilization optimization that are simple, effective and capable of handling unexpected scenarios with minimal underlying assumptions.

The presented approach for efficient resource utilization based on online bin-packing fulfills these requirements in this setting. Our results illustrate efficient scheduling of computing resources based on two very different use cases, where in both cases the framework has limited a priori knowledge of the workload. The presented error plots highlights that despite the noise created by frequently starting and/or stopping the stream processing engines and varying the workload types, HarmonicIO with the IRM can handle this and offers a stable and efficient processing framework. The potential is especially visible in the real use-case experiment, during which the images are streamed at a high frequency allowing the resources to be used consecutively.

As mentioned earlier, HarmonicIO is designed to address the needs for processing large datasets based on relatively large individual objects. It is a specialized streaming framework that is well suited for scientific workflows. The presented solution based on online bin-packing fits well with HarmonicIO and improves the framework to an extent that allows us to compare the framework with the production-grade Spark streaming framework with the recent auto-scaling feature.

In the future, we would like further extend our approach with multi-dimensional online bin-packing. The motivation for this is to be able to profile and schedule workloads based on more resources than only CPU, such as RAM, network usage, or even variations of CPU metrics like average, maximum etc. This would allow us to handle more challenging use cases other than the scientific workflows covered so far.